\documentclass[]{article}
\usepackage{cite}
\usepackage{amsmath,amssymb,amsfonts}
\usepackage{algorithmic}
\usepackage{graphicx}
\usepackage{textcomp}
\usepackage{authblk}
\usepackage{abstract}
\def\BibTeX{{\rm B\kern-.05em{\sc i\kern-.025em b}\kern-.08em
    T\kern-.1667em\lower.7ex\hbox{E}\kern-.125emX}}

\title{GLRT-based Detection in Bistatic Sonar under Strong Direct Blast with Multipath Propagation}

\author[1,2]{Bo Lei \thanks{Corresponding author:lei.bo@nwpu.edu.cn}}
\author[1]{Yao Zhang}
\author[1,2]{Yixin Yang}
\affil[1]{School of Marine Science and Technology, Northwestern Polytechnical University, Xi’an, 710072, China}
\affil[2]{QingDao Reasearch Institute (Northwestern Polytechnical University), QingDao, Shandong, 266200, China}

\begin{document}
\maketitle
\begin{abstract}
Direct blast is a strong interference in bistatic sonar and difficult to suppress due to multipath propagation for blasts and signals. A generalized likelihood ratio test (GLRT) based detection scheme in the frequency domain of the received signals is proposed in this study, and the unknown parameters are estimated using Maximum Likelihood Estimates and Weighted Fourier Transform and Relaxation in a multipath environment. The distributions of the test statistic of detectors for known and unknown noise power are given in theory, and the detection probability is determined. The performance of the detector decreases by 4 dB when the noise power is evaluated with maximum likelihood estimates. Simulations show the effectiveness of the detector under a forward scattering detection configuration with a low signal-to-direct blast ratio. The sensitivity of many factors is discussed, and robustness is achieved.
\newline

\centering
\textbf{keywords:}GLRT, forward scattering detection, bistatic sonar, detection probability.
\end{abstract}

\footnote{This work was supported by the National Natural Science Foundation of China (61571366), the Fundamental Research Funds for the Central Universities (3102018AX003), and the Seed Foundation of Innovation and Creation for Graduate Students in Northwestern Polytechnical University.}

\section{Introduction}
\label{sec:introduction}
Bistatic sonar has many advantages over monostatic sonar; for example, the detection area of bistatic sonar is larger, the concealment is better, and it is not easily suppressed by the underwater acoustic countermeasure device. However, the application of bistatic sonar entails unfavorable factors, and direct blast interference is one of the important problems. When a strong direct blast and the target echo overlap, detection becomes difficult. The aberrations of the received signal caused by the target echo are considerably weak and difficult to detect because the strength of direct blasts is several decibels over the scattered waves when the receiver is far from the target. Furthermore, the dynamics of the environment can cause fluctuations of the received signal, which make the detection of target echoes highly complicated. Overlapping is unavoidable when the target is near the baseline of the bistatic sonar, which leads to a blind zone in bistatic detection using traditional technologies. Short pulse signals have been introduced as a common solution, but they are inefficient underwater because multipath propagation decreases the temporal resolution; thus, weak targets are lost. 
Many researchers have proposed new algorithms and techniques to solve the problem of bistatic sonar detection in blind zones. Gillespie \cite{gillespie1997littoral} and Matveev \cite{matveev2000complex,matveev2007forward} developed intuitive and preliminary methods that use matched filtering techniques to detect weak disturbances caused by the target. Song \textit{et al}. \cite{song2003demonstration} proposed time reversal mirror technology and constructed a high-frequency acoustic barrier. Folegot \textit{et al}. \cite{folegot2008active} developed a localization algorithm on the basis of an ambiguity image composed of sound rays. A premise of this method is that the sound rays pass through the target position, and the target blocks the sound lines. Marandet \textit{et al}. \cite{marandet2011target} proposed a scheme via strength comparison using the double-beamforming technique and designed a laboratory-scale experiment. Sabra \textit{et al}. \cite{sabra2010experimental} extracted the aberrations of received signals on hydrophones caused by a target by using principal component analysis (PCA). Lei \textit{et al}. \cite{lei2014forward} further combined PCA with array processing to enhance the aberrations of the recorded direct arrival data on a short vertical receiver array (VRA), validated the performance of the proposed method via a lake experiment, and identified the relationship between sound field aberration and crossing distance in theory. The adaptive processing scheme and PCA-based concept were executed after acquiring several pulses. He \textit{et al}. \cite{he2015forward} proposed a processing scheme based on adaptive filtering to decrease the degree of received data association, and Lei \textit{et al}. \cite{lei2017detection} developed this technology for a dynamic environment. He \textit{et al}. \cite{he2018direct} developed an approach called low Doppler analysis and recording gram to separate target scattered waves from direct blasts in the Doppler domain. Lei \textit{et al}. \cite{lei2019detection} further proposed the use of an anomaly detection method to detect sound field aberrations caused by the target.
The generalized likelihood ratio test in passive radar signal processing has become an attractive research topic, and modified schemes have been proposed in many studies \cite{kelly1989adaptive,burgess1996subspace,bose1996adaptive,gerlach1999adaptive,gerlach2000fast,abramovich2007modified,xu2015impact}. In \cite{kelly1989adaptive}, \cite{burgess1996subspace}, and \cite{bose1996adaptive}, distributed target detectors in Gaussian white noise and with the constant false alarm rate (CFAR) property were developed based on GLRT. In \cite{gerlach1999adaptive}, Gerlach and Steiner derived a modified GLRT for the adaptive detection of range-spread targets. After outlining the model of the detection problem in the surveillance channel as a composite in the hypothesis test, a GLRT was derived in \cite{gerlach2000fast}. Two detectors \cite{abramovich2007modified} under the GLRT framework were developed for a multistatic passive radar in the presence of noise and direct path interference. Bistatic sonar faces a similar problem as passive bistatic radar, namely, strong interference of direct blast influence and noise. However, serious multipath propagations for the direct blast and scattered signal are present in bistatic sonar, particularly in cases with a very large bistatic angle (e.g., forward scattering detection). In continuous active sonar, a simple generalized likelihood ratio (GLR)-based criterion was developed to enhance the range-Doppler and range-compression images generated by the standard matched filter \cite{xu2015impact}. 
Here we examine the target detection problem for a bistatic sonar with forward scattering wherein the receivers are contaminated by a strong blast. A scheme based on the GLRT framework is proposed to achieve target detection in the blind zone. A hypothesis test model is established based on the received signal models, and model parameters are estimated to derive a generalized likelihood ratio. The first detector assumes that the noise level is known. The noise level in this case must be measured in advance. Another detector is proposed for the case with an unknown noise level. Maximum likelihood (ML) in the closed form and weighted Fourier transform and relaxation (WRELAX) algorithm \cite{li1998efficient} were used for the parameter estimates. Numerical results are provided to show the performance of the proposed detectors in forward scattering detection.
The remainder of this paper is organized as follows. Section \uppercase\expandafter{\romannumeral2} is devoted to the problems in bistatic sonar. Section \uppercase\expandafter{\romannumeral3} presents a hypothesis test and describes the ML estimate of the model parameters. Section \uppercase\expandafter{\romannumeral4} shows the derivation of the generalized likelihood ratio in the two cases, and Section \uppercase\expandafter{\romannumeral5} analyzes the performance of GLRT. The conclusions are provided in Section \uppercase\expandafter{\romannumeral6}.

Notation: Vectors (matrices) are denoted by boldface lower (upper) case letters, and all vectors are column ones. Superscripts ${{(\cdot )}^{*}}$, ${{(\cdot )}^{T}}$, and ${{(\cdot )}^{H}}$ denote complex conjugate, transpose, and complex conjugate transpose, respectively. The Euclidean norm of vector $\mathbf{x}$ is denoted by $\left\| \mathbf{x} \right\|$, the Frobenius norm of vector $\mathbf{x}$ is denoted by $\left\| \mathbf{x} \right\|_{\mathrm{F}}$, and $\left| x \right|$ represents the modulus of $x$. $\mathbf{I}$ denotes the identity matrix, and $\operatorname{tr}(\mathbf{A})$ is the trace of square matrix $\mathbf{A}$.

\section{PROBLEMS IN BISTATIC SONAR}
\begin{figure}[ht]
	
	\centering
	\includegraphics[scale=0.4]{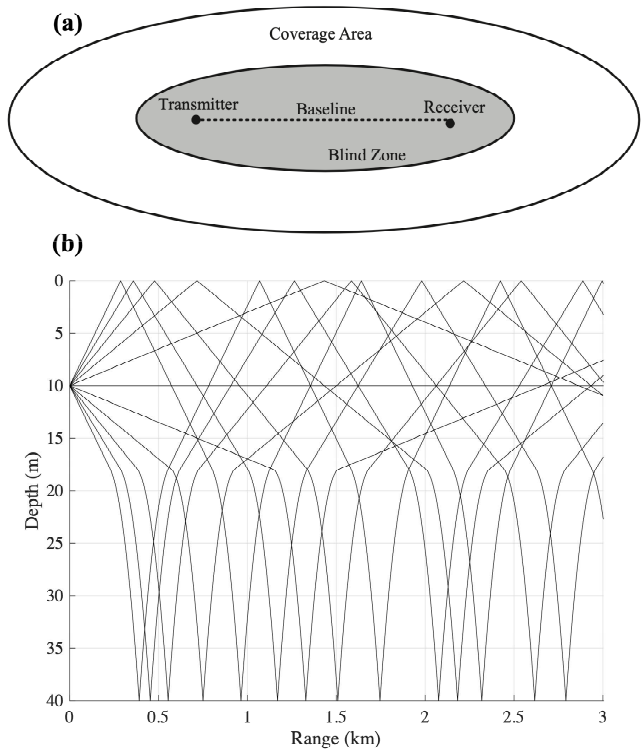}
	\caption{Schematic of bistatic sonar under a multipath environment. (a) Coverage area and blind zone of bistatic sonar detection with traditional technology due to the strong direct blast. (b) Multipath propagation in the shallow water environment.}
	\label{fig1}
\end{figure}

Consider a shallow water multipath environment where the transmitter and receiver are separated for a bistatic sonar system (Fig. \ref{fig1}). A priori known signal $s(t)$ is transmitted repeatedly to the receiver through a multipath environment. When the target is absent, the received signal is called a direct blast. It is expressed as
\begin{equation}\label{1}
{{x}_{0}}(t)=\sum\limits_{i=1}^{M}{{{a}_{i}}}s\left( t-{{\tau }_{i}^\mathrm{d}} \right)+n(t),
\end{equation}
where $M$ is the number of multipath components. $\left\{ {{a_i}} \right\}_{i = 1}^M$ and $\left\{ {{\tau _i^\mathrm{d}}} \right\}_{i = 1}^M$ are the complex path factor and time delay for each multipath, respectively. $n(t)$ is complex white Gaussian noise and $n(t)\sim CN(0,{{\sigma }^{2}})$. The target scattered signals propagate through a multipath channel and are hence covered by the direct blast. The mixture can be expressed as
\begin{equation}\label{2}
{x_1}(t) = \sum\limits_{i = 1}^M {{a_i}} s\left( {t - {\tau _i^\mathrm{d}}} \right) + \sum\limits_{k = 1}^K {{b_k}} s\left( {t - {\tau _k^\mathrm{s}}} \right){e^{{\rm{j}}2\pi {f_{\rm{d}}}t}} + n(t),
\end{equation}
where $K$ is the number of multipath components for scattered signals. $\left\{ {{b}_{k}} \right\}_{k=1}^{K}$ and $\left\{ {{\tau }_{k}^\mathrm{s}} \right\}_{k=1}^{K}$ are the complex path factor and time delay for each multipath of the scattered signal, respectively. ${{f}_{\text{d}}}$ is the Doppler shift. This component is extremely weak for the case of detection in a blind zone.

Considering the frequency domain form, \eqref{1} and \eqref{2} are subjected to fast Fourier transform (FFT), thereby yielding
\begin{equation}
\begin{aligned}
{{\mathbf{X}}_0} &= {{\mathbf{\Phi }}_{\mathrm{d}}}({\mathbf{\tau^d }}){\mathbf{a}} + {\mathbf{W}},\\
{{\mathbf{X}}_1} &= {{\mathbf{\Phi }}_{\mathrm{d}}}({\mathbf{\tau^d }}){\mathbf{a}} + {{\mathbf{\Phi }}_{\mathrm{s}}}({\mathbf{\tau^s }}){\mathbf{b}} + {\mathbf{W}},
\end{aligned}
\end{equation}
where $\mathbf{X}={{[\mathbf{X}(0),\mathbf{X}(1),\ldots ,\mathbf{X}(N-1)]}^{T}}$ is viewed as the received vector and$\mathbf{W}={{[\mathbf{W}(0),\mathbf{W}(1),\ldots ,\mathbf{W}(N-1)]}^{T}}$is a complex white Gaussian noise vector $n(t)$ containing the FFT coefficients of samples of the additive noise; thus, $\mathbf{W}\sim CN(0,N{{\sigma }^{2}})$. $N$ is the number of FFT points. $\mathbf{a}=\left[a_{1}, a_{2}, \dots, a_{M}\right]^{T}$and $\mathbf{b}=\left[b_{1}, b_{2}, \dots, b_{K}\right]^{T}$. Matrices ${{\mathbf{\Phi} }_{\mathrm{d}}}(\tau )$ and ${{\mathbf{\Phi} }_{\mathrm{s}}}(\tau )$ depend only on the unknown delays of the multipath direct blast and scattered signal, respectively. They are given by
\begin{equation}
\begin{aligned}
{{\mathbf{\Phi }}_{\mathrm{d}}}({\mathbf{\tau^\mathrm{d} }}) = \left[ {\phi \left( {{\tau _1^\mathrm{d}}} \right),\phi \left( {{\tau _2^\mathrm{d}}} \right), \ldots ,\phi \left( {{\tau _M^\mathrm{d}}} \right)} \right],\\
{{\mathbf{\Phi }}_{\mathrm{s}}}({\mathbf{\tau^\mathrm{s} }}) = \left[ {\phi \left( {{\tau _1^\mathrm{s}}} \right),\;\phi \left( {{\tau _2^\mathrm{s}}} \right), \ldots ,\phi \left( {{\tau _K^\mathrm{s}}} \right)} \right],
\end{aligned}
\end{equation}
with the columns defined as
\begin{equation}
\resizebox{.95\hsize}{!}{$\phi \left( {{\tau _i}} \right) = {\left[ {\mathbf{S}(0),\mathbf{S}(1){e^{ - \frac{{j2\pi {f_s}{\tau _i}}}{N}}}, \ldots ,\mathbf{S}(N - 1){e^{ - \frac{{j2\pi {f_s}(N - 1){\tau _i}}}{N}}}} \right]^{T},}$}
\end{equation}
where $\mathbf{S}={{[\mathbf{S}(0),\mathbf{S}(1),\ldots ,\mathbf{S}(N-1)]}^{T}}$is viewed as the transmitted vector and ${{f}_{s}}$ is the sampling frequency.

The general problem of detecting a target characterized by unknown complex amplitude $\bf{b}$, unknown time delay $\left[ {{\tau }_{1}},\ldots ,{{\tau }_{K}} \right]$, and Doppler frequency $f_d$ in the presence of clutter/multipath with unknown complex amplitude $\bf{a}$ and unknown time delay $\left[ {{\tau }_{1}},\ldots ,{{\tau }_{M}} \right]$ can be described by the hypotheses
\begin{equation}
\begin{aligned}
{{\cal H}_0}:{\mathbf{X}} &= {{\mathbf{\Phi }}_{\mathrm{d}}}({\mathbf{\tau^\mathrm{d} }}){\mathbf{a}} + {\mathbf{W}},\\
{{\cal H}_1}:{\mathbf{X}} &= {{\mathbf{\Phi }}_{\mathrm{d}}}({\mathbf{\tau^\mathrm{d} }}){\mathbf{a}} + {{\mathbf{\Phi }}_{\mathrm{s}}}({\mathbf{\tau^\mathrm{s} }}){\mathbf{b}} + {\mathbf{W}}.
\end{aligned}
\end{equation}

To solve the detection problem, the following assumptions are made:
\begin{enumerate}
	\item The number of multipaths for direct blast $M$ and scattered signal $K$ is assumed to be unknown.
	\item The direct blast amplitudes of multipath complex-valued $\left[ {{a}_{1}},\ldots ,{{a}_{M}} \right]$ and their delays $\left[ \tau _{1}^{\mathrm{d}},\ldots ,\tau _{M}^{\mathrm{d}} \right]$ are assumed to be deterministic and unknown. The scattered signal amplitudes of multipath complex-valued $\left[ {{b}_{1}},\ldots ,{{b}_{M}} \right]$ and their delays $\left[ \tau _{1}^{\mathrm{s}},\ldots ,\tau _{M}^{\mathrm{s}} \right]$ are assumed to be deterministic and unknown. 
	\item Doppler shift is ignored in the hypothesis for two reasons. First, the underwater vehicle is generally at a low speed, and the transmitted signal works at a low frequency. Hence, the Doppler shift is not obvious. Second, the Doppler shift of the target in the bistatic sonar configuration can be expressed as
	\begin{equation}
	{f_{\rm{d}}} = \frac{{2vf}}{c}\cos \frac{\beta }{2}\cos \theta ,
	\end{equation}
	where $f$ is the transmitted signal frequency, $v$ is the target speed, $c$ is the sound speed, $\beta$ is the bistatic separation angle, and $\theta$ is the angle between the target motion direction and angle bisector of $\beta$. In this scenario, when the target is in the detection blind zone near the baseline, $\beta \approx \pi $, and the Doppler shift is very weak.
	\item The two cases of known and unknown spectrum power of the noise ${{\sigma }^{2}}$ are discussed separately.
\end{enumerate}

The probability density function (PDF) of $\mathbf{X}$ parameterized by $\mathbf{a}$ and $\tau^\mathrm{d}$ can be written as
\begin{equation}
f\left( {{\mathbf{X}}|{{\cal H}_0}} \right) = \frac{1}{{{{\left( {2\pi N{\sigma ^2}} \right)}^{N/2}}}}\exp \left\{ { - \frac{{{{\left[ {{\mathbf{X}} - {{\mathbf{\Phi }}_{\mathrm{d}}}{\mathbf{a}}} \right]}^{H}}\left[ {{\mathbf{X}} - {{\mathbf{\Phi }}_{\mathrm{d}}}{\mathbf{a}}} \right]}}{{2N{\sigma ^2}}}} \right\}
\end{equation}
under ${{\cal{H}}_{0}}$. The PDF of $\mathbf{X}$ parameterized by $\mathbf{a}$, $\mathbf{b}$, $\tau^d$, and $\tau^s$ can be written as
\begin{equation}\label{9}
\begin{aligned}
f\left( {{\mathbf{X}}|{{\cal H}_{\mathrm{1}}}} \right) &= \frac{1}{{{{\left( {2\pi N{\sigma ^2}} \right)}^{N/2}}}}\cdot\\
&\exp \left\{ { - \frac{{{{\left[ {{\mathbf{X}} - {{\mathbf{\Phi }}_{\mathrm{d}}}{\mathbf{a}} - {{\mathbf{\Phi }}_{\mathrm{s}}}{\mathbf{b}}} \right]}^{H}}\left[ {{\mathbf{X}} - {{\mathbf{\Phi }}_{\mathrm{d}}}{\mathbf{a}} - {{\mathbf{\Phi }}_{\mathrm{s}}}{\mathbf{b}}} \right]}}{{2N{\sigma ^2}}}} \right\}
\end{aligned}
\end{equation}
under ${{\cal{H}}_{1}}$. 

According to the Newman–Pearson criterion \cite{neyman1933ix}, the optimal solution to the hypothesis testing problem is the likelihood ratio test, but it cannot be directly implemented here because knowledge of the parameters $\mathbf{a}$,$\mathbf{b}$, $\left\{ \tau _{i}^{\mathrm{d}}\right\}_{i=1}^{M}$, and $\left\{ \tau _{k}^{\mathrm{s}} \right\}_{k=1}^{K}$ are required; these parameters are unknown in practical situations. Their maximum likelihood estimates (MLEs) in the likelihood ratio test provide good solutions.

\section{Parameters Estimation}
\subsection{Maximum likelihood estimates of path factors}
When the target is not present, the received signal is the same as the direct blast, and the path factor can be estimated as
\begin{equation}
{\widehat {\mathbf{a}}_0} = {\left( {{\mathbf{\Phi }}_{\mathrm{d}}^{H}{{\mathbf{\Phi }}_{\mathrm{d}}}} \right)^{ - 1}}{\mathbf{\Phi }}_{\mathrm{d}}^{H}{\mathbf{X}}
\end{equation}
under hypothesis ${\mathcal{H}}_{0}$. Compared with hypothesis ${\mathcal{H}}_{0}$, estimating the path factors under hypothesis ${\mathcal{H}}_{1}$ is complicated. For the convenience of expression, we define
\begin{equation}
{\mathbf{Q}} = \left[ {{{\mathbf{\Phi }}_{\mathrm{d}}},{{\mathbf{\Phi }}_{\mathrm{s}}}} \right],
\end{equation}
and
\begin{equation}
\mathbf{c} = \left[\mathbf{a},\mathbf{b}\right]^T.
\end{equation}
Then, \eqref{9} can be simplified to
\begin{equation}
f\left( {{\mathbf{X}}|{{\cal H}_{\mathrm{1}}}} \right) = \frac{1}{{{{\left( {2\pi N{\sigma ^2}} \right)}^{N/2}}}}\exp \left\{ { - \frac{{{{[{\mathbf{X}} - {\mathbf{Qc}}]}^{H}}[{\mathbf{X}} - {\mathbf{Qc}}]}}{{2N{\sigma ^2}}}} \right\}.
\end{equation}

With Lemma 1, path factors $\mathbf{c}$ can be estimated as \eqref{14} and \eqref{15}.
\begin{figure*}[!t]
	\normalsize
	\setcounter{equation}{13}
	\begin{equation}\label{14}
	\begin{aligned}
	{\widehat{\mathbf{c}}_1}&={\left({{{\mathbf{Q}}^{H}}{\mathbf{Q}}}\right)^{-1}}{{\mathbf{Q}}^{H}}{\mathbf{x}} \\
	&= {\left[ {\begin{array}{*{20}{c}}{\mathbf{\Phi} _{\mathrm{d}}^{H}{\mathbf{\Phi} _{\mathrm{d}}}}&{\mathbf{\Phi} _{\mathrm{d}}^{H}{\mathbf{\Phi} _{\mathrm{s}}}}\\{\mathbf{\Phi} _{\mathrm{s}}^{H}{\mathbf{\Phi} _{\mathrm{d}}}}&{\mathbf{\Phi} _{\mathrm{s}}^{H}{\mathbf{\Phi} _{\mathrm{s}}}}\end{array}}\right]^{ - 1}}\left[ {\begin{array}{*{20}{c}}{\mathbf{\Phi} _{\mathrm{d}}^{H}{\mathbf{X}}}\\
	{\mathbf{\Phi} _{\mathrm{s}}^{H}{\mathbf{X}}}\end{array}} \right] \\
	&= \left[ {\begin{array}{*{20}{c}}{{{\left( {\mathbf{\Phi} _{\mathrm{d}}^{H}{\mathbf{\Phi} _{\mathrm{d}}}} \right)}^{-1}} \!+\! {{\left( {\mathbf{\Phi} _{\mathrm{d}}^{H}{\mathbf{\Phi} _{\mathrm{d}}}} \right)}^{ - 1}}\mathbf{\Phi} _{\mathrm{d}}^{{H}}{\mathbf{\Phi} _{\mathrm{s}}}{\bf{K}}\mathbf{\Phi} _s^{H}{\mathbf{\Phi} _{\mathrm{d}}}{{\left( {\mathbf{\Phi} _{\mathrm{d}}^{{H}}{\mathbf{\Phi} _{\mathrm{d}}}} \right)}^{ - 1}}}&{ - {{\left( {\mathbf{\Phi} _{\mathrm{d}}^{{H}}{\mathbf{\Phi} _{\mathrm{d}}}} \right)}^{ - 1}}\mathbf{\Phi} _{\mathrm{d}}^{{H}}{\mathbf{\Phi} _{\mathrm{s}}}{\mathbf{K}}}\\{ - {\mathbf{K}}\mathbf{\Phi} _{\mathrm{s}}^{{H}}{\mathbf{\Phi} _{\mathrm{d}}}{{\left( {\mathbf{\Phi} _{\mathrm{d}}^{{H}}{\mathbf{\Phi} _{\mathrm{d}}}} \right)}^{ - 1}}}&{\mathbf{K}}
	\end{array}} \right]\left[ {\begin{array}{*{20}{c}}{\mathbf{\Phi} _{\mathrm{d}}^{{H}}{\mathbf{X}}}\\{\mathbf{\Phi} _{\mathrm{s}}^{{H}}{\mathbf{X}}}\end{array}} \right]\\
	&= \left[ {\begin{array}{*{20}{c}}
	{{{\widehat {\mathbf{a}}}_0} + {{\left( {\mathbf{\Phi} _{\mathrm{d}}^{{H}}{\mathbf{\Phi} _{\mathrm{d}}}} \right)}^{ - 1}}\mathbf{\Phi} _{\mathrm{d}}^{{H}}{\mathbf{\Phi} _{\mathrm{s}}}{\mathbf{K}}\mathbf{\Phi} _{\mathrm{s}}^{{H}}\left( {{\mathbf{\Phi} _{\mathrm{d}}}{{\widehat {\mathbf{a}}}_0} - {\mathbf{X}}} \right)}\\
	{ - {\mathbf{K}}\mathbf{\Phi} _{\mathrm{s}}^{{H}}\left( {{\mathbf{\Phi} _{\mathrm{d}}}{{\widehat {\mathbf{a}}}_0} - {\mathbf{X}}} \right)}
	\end{array}} \right],
	\end{aligned}
	\end{equation}
	
	\begin{equation}\label{15}
	{\mathbf{K}} = {\left( {\mathbf{\Phi} _\mathrm{s}^{{H}}{\mathbf{\Phi} _{\mathrm{s}}} - \mathbf{\Phi} _{\mathrm{s}}^{{H}}{\mathbf{\Phi} _{\mathrm{d}}}{{\left( {\mathbf{\Phi} _{\mathrm{d}}^{{H}}{\mathbf{\Phi} _{\mathrm{d}}}} \right)}^{ - 1}}\mathbf{\Phi} _{\mathrm{d}}^{{H}}{\mathbf{\Phi} _{\mathrm{s}}}} \right)^{ - 1}}
	= {\left[ {\mathbf{\Phi} _{\mathrm{s}}^{{H}}\left( {{\mathbf{I}} - {\mathbf{\Phi} _{\mathrm{d}}}{{\left( {\mathbf{\Phi} _{\mathrm{d}}^{{H}}{\mathbf{\Phi} _{\mathrm{d}}}} \right)}^{ - 1}}\mathbf{\Phi} _{\mathrm{d}}^{{H}}} \right){\mathbf{\Phi} _s}} \right]^{ - 1}}.
	\end{equation}
	\setcounter{equation}{15}
	\hrulefill
	\vspace * {4pt}
\end{figure*}
Then we derive the estimates
\begin{equation}
\left[ {\begin{array}{*{20}{c}}
	{{{\widehat {\mathbf{a}}}_1}}\\
	{{{\widehat {\mathbf{b}}}_1}}
	\end{array}} \right] = \left[ {\begin{array}{*{20}{c}}
	{{{\widehat {\bf{a}}}_0} + {{\left( {{\mathbf{\Phi }}_{\mathrm{d}}^{{H}}{{\mathbf{\Phi }}_{\mathrm{d}}}} \right)}^{ - 1}}{\mathbf{\Phi }}_{\mathrm{d}}^{{H}}{{\mathbf{\Phi }}_{\mathrm{s}}}{\mathbf{K\Phi }}_{\mathrm{s}}^{{H}}\left( {{{\mathbf{\Phi }}_{\mathrm{d}}}{{\widehat {\mathbf{a}}}_0} - {\mathbf{X}}} \right)}\\
	{ - {\mathbf{K\Phi }}_{\mathrm{s}}^{{H}}\left( {{{\mathbf{\Phi }}_{\mathrm{d}}}{{\widehat {\mathbf{a}}}_0} - {\mathbf{X}}} \right)}
	\end{array}} \right].
\end{equation}

\subsection{Estimates of parameters of time delays}
Time delay estimation is a classic problem in sonar signal processing. Researchers have proposed several time delay estimators, such as matched filter approach \cite{ehrenberg1978signal}, expectation maximization (EM) algorithm \cite{feder1988parameter}, and WRELAX algorithm \cite{li1998efficient}. Compared with other existing algorithms, WRELAX is more systematic and efficient and has fewer limitations on the signal waveforms. Furthermore, the mean squared error (MSE) of WRELAX is very close to the corresponding Cramer-Rao lower bound (CRLB) for a wide range of signal-to-noise ratios (SNRs). Therefore, the WRELAX algorithm is applied here to estimate $\tau ^{\mathrm{d}}$ and $\tau ^{\mathrm{s}}$.

The processing steps of the WRELAX algorithm are as follows:
\begin{enumerate}
	\item Assume that the number of multipaths is $M = 1$. Compute ${{a}_{1}},{{\tau }_{1}}$.
	\item Assume that the number of multipaths is $M = 2$. Compute ${{a}_{2}},{{\tau }_{2}}$ from the residual signal and redetermine ${{a}_{1}},{{\tau }_{1}}$.
	
	Repeat the two previous sub-steps until convergence is achieved.
	\item Assume that the number of multipaths is $M = 3$. Compute ${{a}_{3}},{{\tau }_{3}}$ from the residual signal. Redetermine ${{a}_{1}},{{\tau }_{1}}$ then ${{a}_{2}},{{\tau }_{2}}$.
	
	Repeat the three previous sub-steps until convergence is achieved.
	\item Continue similarly until $M$ is equal to the desired or estimated number of signals. 
\end{enumerate}

The residual signal and estimators are as follows:
\begin{equation}
{{\mathbf{X}}_m} = {\mathbf{X}} - \mathop \sum \limits_{i = 1,i \ne m}^M {\hat a_i}\phi \left( {{{\hat \tau }_i}} \right),
\end{equation}
\begin{equation}
{\hat \tau _m} = \arg {\max _{{\tau _m}}}\left| {\phi \left( {{\tau _m}} \right)*{{\mathbf{X}}_m}} \right|,
\end{equation}
\begin{equation}
{\hat a_m} = {\left. {\frac{{\phi \left( {{\tau _m}} \right)*{{\mathbf{X}}_m}}}{{\left\| {\mathbf{S}} \right\|_{\mathrm{F}}^2}}} \right|_{{\tau _m} = {{\hat \tau }_m}}}.
\end{equation}

To estimate the multipath delay, we need to determine the number of multipaths. Therefore, some prior information is needed to decide the number of multipaths. The maximum number of multipaths can be evaluated with the underwater sound propagation model. The evaluated path number might be larger than that in the real environment because sounds with large grazing angles attenuate rapidly. The evaluation of these extra but non-existent paths may not result in the performance decrease, since it provides a maximum likelihood estimates. This intuition will be discussed in the simulation section. Furthermore, the number of multipaths of the scattered signal is larger than that of the direct blast, but many of them have only small path factors that can be ignored. For convenience, we assume that $M = K$. Therefore, M delays ${{\mathbf{\hat{\tau }}}^{\text{d}}}=[{{\tau }_{1}},{{\tau }_{2}},\cdots ,{{\tau }_{M}}]$ are estimated for hypothesis ${{\mathcal{H}}_{0}}$, and 2M delays ${{\mathbf{\hat{\tau }}}^{\text{d}}}\cup {{\mathbf{\hat{\tau }}}^{\text{s}}}=[{{\tau }_{1}},{{\tau }_{2}},\cdots ,{{\tau }_{2M}}]$ are estimated for hypothesis ${{\mathcal{H}}_{1}}$. Then, the $M$ time delay estimates that are the closest to $\hat{\tau} ^{\rm{d}}$ are deleted, and the remaining $M$ time delay estimates are $\hat{\tau} ^{\rm{s}}$. 

\subsection{Maximum likelihood estimates of spectrum power of noise}
In practice, noise strength is usually known because it has a considerable impact on sonar performance. However, noise might vary due to shipping, winds, topography, and other factors. Hence, it is difficult to know well. If noise parameter   is unknown, then it is estimated as
\begin{equation}
\sigma _0^2 = \frac{1}{{{N^2}}}{[{\mathbf{X}} - {{\mathbf{\Phi }}_{\mathrm{d}}}{\widehat{\mathbf{a}}}]^{{H}}}[{\mathbf{X}} - {{\mathbf{\Phi }}_{\mathrm{d}}}{\widehat{\mathbf{a}}}]
\end{equation}
under hypothesis ${\mathcal{H}}_{0}$ and as
\begin{equation}
\sigma _1^2 = \frac{1}{{{N^2}}}{[{\mathbf{X}} - {{\mathbf{\Phi }}_{\mathrm{d}}}\widehat{\mathbf{a}} - {{\mathbf{\Phi }}_{\mathrm{s}}}\widehat{\mathbf{b}}]^{{H}}}[{\mathbf{X}} - {{\mathbf{\Phi }}_{\mathrm{d}}}{\widehat{\mathbf{a}}} - {{\mathbf{\Phi }}_{\mathrm{s}}}\widehat{\mathbf{b}}]
\end{equation}
under hypothesis ${{\mathcal{H}}_{1}}$.

\section{GLRT DERIVATIONS}
\subsection{Known spectrum power of noise}
GLRT can be written as
\begin{equation}
{L_{{\mathrm{GLR}}}}({\mathbf{X}}) = \frac{{{{\max }_{{\mathbf{a}},{\mathbf{b}},{{\mathbf{\tau }}^d},{{\mathbf{\tau }}^{\mathbf{s}}}}}f\left( {{\mathbf{X}}|{\mathbf{a}},{\mathbf{b}},{{\mathbf{\tau }}^{\mathrm{d}}},{{\mathbf{\tau }}^{\mathrm{s}}};{{\cal H}_1}} \right)}}{{{{\max }_{{\mathbf{a}},{{\mathbf{\tau }}^{\mathrm{d}}}}}f\left( {{\mathbf{X}}|{\mathbf{a}},{{\mathbf{\tau }}^{\mathrm{d}}};{{\cal H}_0}} \right)}}
\mathop{\gtrless}\limits_{{\cal H}_0}^{{\cal H}_1} \eta .
\end{equation}

Substituting the maximum likelihood estimates (MLEs) of the parameters under each hypothesis into the PDFs, constructing the likelihood ratio, taking the logarithm of it, and performing some simplification yield the test statistic
\begin{equation}
\begin{split}
\begin{aligned}
&\ln {l_{{\text{GLR}}}}({\mathbf{X}})  \\
&= - \frac{1}{{2N{\sigma ^2}}}
\left\{ {{{\left[ {{\mathbf{X}} - {{\mathbf{\Phi }}_{\mathrm{d}}}{{\widehat {\mathbf{a}}}_{\mathrm{1}}} - {{\mathbf{\Phi }}_{\mathrm{s}}}{{\widehat {\mathbf{b}}}_{\mathrm{1}}}} \right]}^{{H}}}\left[ {{\mathbf{X}} - {{\mathbf{\Phi }}_{\mathrm{d}}}{{\widehat {\mathbf{a}}}_1} - {{\mathbf{\Phi }}_{\mathrm{s}}}{{\widehat {\mathbf{b}}}_1}} \right]} \right.\\
&\left. { - {{\left[ {{\mathbf{X}} - {{\mathbf{\Phi }}_{\mathrm{d}}}{{\widehat {\mathbf{a}}}_0}} \right]}^{{H}}}\left[ {{\mathbf{X}} - {{\mathbf{\Phi }}_{\mathrm{d}}}{{\widehat {\mathbf{a}}}_0}} \right]} \right\}\\
& = \frac{{{{\mathbf{X}}^{{H}}}\left[ {{\mathbf{Q}}{{\left( {{{\mathbf{Q}}^{{H}}}{\mathbf{Q}}} \right)}^{ - 1}}{{\mathbf{Q}}^{{H}}} - {{\mathbf{\Phi }}_{\mathrm{d}}}{{\left( {{\mathbf{\Phi }}_{\mathrm{d}}^{{H}}{{\mathbf{\Phi }}_{\mathrm{d}}}} \right)}^{ - 1}}{\mathbf{\Phi }}_{\mathrm{d}}^{{H}}} \right]{\mathbf{X}}}}{{2N{\sigma ^2}}}\\
& = \frac{1}{{2N{\sigma ^2}}}\left\{ {{{\mathbf{X}}^{H}}\left( {{\mathbf{I}} - {{\mathbf{\Phi }}_{\mathrm{d}}}{{\left( {{\mathbf{\Phi }}_{\mathrm{d}}^{{H}}{{\mathbf{\Phi }}_{\mathrm{d}}}} \right)}^{ - 1}}{\mathbf{\Phi }}_{\mathrm{d}}^{{H}}} \right){{\mathbf{\Phi }}_{\mathrm{s}}}{\mathbf{K\Phi }}_{\mathrm{s}}^{{H}}} \right.\\
& \cdot \left. {\left( {{\mathbf{I}} - {{\mathbf{\Phi }}_{\mathrm{d}}}{{\left( {{\mathbf{\Phi }}_{\mathrm{d}}^{{H}}{{\mathbf{\Phi }}_{\mathrm{d}}}} \right)}^{ - 1}}{\mathbf{\Phi }}_{\mathrm{d}}^{{H}}} \right){\mathbf{X}}} \right\}
\mathop{\gtrless}\limits_{{\cal H}_0}^{{\cal H}_1} \eta.
\end{aligned}
\end{split}
\end{equation}

When only one path exists for the signal, matrix $\mathbf{K}$ decreases to one dimension, and the expression is similar as (30) in \cite{zaimbashi2013glrt}. The test statistic can be further expressed as 
\begin{equation}\label{24}
\begin{aligned}
{T_0}({\mathbf{X}}) &= 2\ln {l_{{\mathrm{GLR}}}}({\mathbf{X}})\\
&= \frac{1}{{N{\sigma ^2}}}{{\mathbf{X}}^{{H}}}\left( {{\mathbf{I}} - {{\mathbf{\Phi }}_{\mathrm{d}}}{{\left( {{\mathbf{\Phi }}_{\mathrm{d}}^{{H}}{{\mathbf{\Phi }}_{\mathrm{d}}}} \right)}^{ - 1}}{\mathbf{\Phi }}_{\mathrm{d}}^{{H}}} \right){{\mathbf{\Phi }}_{\mathrm{s}}}\\
&\cdot {\left[ {{\mathbf{\Phi }}_s^{{H}}\left( {{\mathbf{I}} - {{\mathbf{\Phi }}_{\mathrm{d}}}{{\left( {{\mathbf{\Phi }}_{\mathrm{d}}^{{H}}{{\mathbf{\Phi }}_{\mathrm{d}}}} \right)}^{ - 1}}{\mathbf{\Phi }}_{\mathrm{d}}^{{H}}} \right){{\mathbf{\Phi }}_{\mathrm{s}}}} \right]^{ - 1}} \\
&\cdot {\mathbf{\Phi }}_{\mathrm{s}}^{{H}}\left( {{\mathbf{I}} - {{\mathbf{\Phi }}_{\mathrm{d}}}{{\left( {{\mathbf{\Phi }}_{\mathrm{d}}^{{H}}{{\mathbf{\Phi }}_{\mathrm{d}}}} \right)}^{ - 1}}{\mathbf{\Phi }}_{\mathrm{d}}^{{H}}} \right){\mathbf{X}}\\
&= \frac{1}{{N{\sigma ^2}}}{{\mathbf{X}}^{{H}}}{\mathbf{P}}_{\mathrm{d}}^ \bot {{\mathbf{\Phi }}_{\mathrm{s}}}{\left[ {{\mathbf{\Phi }}_{\mathrm{s}}^{{H}}{\mathbf{P}}_{\mathrm{d}}^ \bot {{\mathbf{\Phi }}_{\mathrm{s}}}} \right]^{ - 1}}{\mathbf{\Phi }}_{\mathrm{s}}^{{H}}{\mathbf{P}}_{\mathrm{d}}^ \bot {\mathbf{X}}\\
&= \frac{1}{{N{\sigma ^2}}}{\left\| {{{\mathbf{\Lambda }}^{ - \frac{1}{2}}}{{\mathbf{U}}^{{H}}}{\mathbf{\Phi }}_{\rm{s}}^{{H}}{\mathbf{P}}_{\mathrm{d}}^ \bot {\mathbf{X}}} \right\|^2}
\mathop{\gtrless}\limits_{{\cal H}_0}^{{\cal H}_1} \eta,
\end{aligned}
\end{equation}
where ${{\mathbf{P}}_{\text{d}}}={\mathbf{\Phi }_{\text{d}}}{{\left( \mathbf{\Phi} _{\text{d}}^{{H}}{{\mathbf{\Phi} }_{\text{d}}} \right)}^{-1}}\mathbf{\Phi} _{\text{d}}^{{H}}$ is the projection matrix that projects a vector onto the column of $\mathbf{\Phi }_{\text{d}}$ and $\mathbf{P}_{\text{d}}^{\bot }=\mathbf{I}-{\mathbf{\Phi }_{\text{d}}}{{\left( \mathbf{\Phi} _{\text{d}}^{{H}}{{\mathbf{\Phi} }_{\text{d}}} \right)}^{-1}}\mathbf{\Phi} _{\text{d}}^{{H}}$ is the orthogonal projection matrix that projects a vector onto the space orthogonal to that spanned by the columns of $\mathbf{\Phi}_\text{d}$. The matrix $\mathbf{\Phi} _{\text{s}}^{{H}}\mathbf{P}_{\text{d}}^{\bot }{\mathbf{\Phi}_{\text{s}}}=\mathbf{U}\Lambda {{\mathbf{U}}^{{H}}}$ involves eigen decomposition. Orthogonal projection matrix $\mathbf{P_\text{d}^\perp}$ has the Hermitian and idempotent property, that is,$\mathbf{P}_{\text{d}}^{\bot }={{\left( \mathbf{P}_{\text{d}}^{\bot } \right)}^{{H}}}$ and $\mathbf{P}_{\text{d}}^{\perp}={{\left( \mathbf{P}_{\text{d}}^{\perp} \right)}^{2}}$.

If the subspace spanned by the columns of $\mathbf{\Phi }_{\text{d}}$ is considered a blast subspace and the orthogonal subspace is regarded as the signal subspace, $\mathbf{P}_{\text{d}}^{\perp}$ in \eqref{24} represents the projection of received vector $\mathbf{X}$ onto the signal subspace at the given time delays of the direct blast. When the threshold of detection is independent of the blast power, the detection exhibits the CFAR property. Equation \eqref{24} indicates that GLRT is dependent on the signal subspace dimension or number of components. Ref. \cite{scharf1991statistical} showed that given matrix $\mathbf{A}$, the quadratic form $\mathbf{X}^{{H}}\mathbf{AX}$ is $\chi^{2}$ distributed with $2p$ degrees of freedom, and $p$ is the rank of $\mathbf{A}$ if and only if $\mathbf{A}$ is idempotent, i.e., $\mathbf{A}^2=\mathbf{A}$. With this property, the detection statistic is $\chi^{2}$ distributed with $2p$ degrees of freedom. Consequently, 
\begin{equation}\label{25}
\begin{aligned}
\frac{{{{\mathbf{X}}^{{H}}}}}{{\sqrt N \sigma }}{\mathbf{P}}_{\mathrm{d}}^ \bot {{\mathbf{\Phi }}_{\mathrm{s}}}{\left[ {{\mathbf{\Phi }}_{\mathrm{s}}^{{H}}{\mathbf{P}}_{\mathrm{d}}^ \bot {{\mathbf{\Phi }}_{\mathrm{s}}}} \right]^{ - 1}}{\mathbf{\Phi }}_{\mathrm{s}}^{{H}}{\mathbf{P}}_{\mathrm{d}}^ \bot \frac{{\mathbf{X}}}{{\sqrt N \sigma }} \\
\sim\left\{ {\begin{array}{*{20}{c}}
	{\chi _v^2({\delta _0})}&{{\text{under }}{{\cal H}_0}}\\
	{\chi _v^2({\delta _1})}&{{\text{under }}{{\cal H}_1}}
	\end{array}} \right. .
\end{aligned}
\end{equation}

Hence, we have
\begin{equation}
\begin{aligned}
v &= {\operatorname{rank}}\left\{ {{\mathbf{P}}_{\mathrm{d}}^ \bot {{\mathbf{\Phi }}_{\mathrm{s}}}{{\left[ {{\mathbf{\Phi }}_{\mathrm{s}}^{{H}}{\mathbf{P}}_{\mathrm{d}}^ \bot {{\mathbf{\Phi }}_{\mathrm{s}}}} \right]}^{ - 1}}{\mathbf{\Phi }}_{\mathrm{s}}^{{H}}{\mathbf{P}}_{\mathrm{d}}^ \bot } \right\}\\
&\le \min \left\{ {{\operatorname{rank}}\left\{ {{\mathbf{P}}_{\mathrm{d}}^ \bot {{\mathbf{\Phi }}_{\mathrm{s}}}{{\left[ {{\mathbf{\Phi }}_{\mathrm{s}}^{{H}}{\mathbf{P}}_{\mathrm{d}}^ \bot {{\mathbf{\Phi }}_{\mathrm{s}}}} \right]}^{ - 1}}} \right\},{\operatorname{rank}}\left( {{\mathbf{\Phi }}_{\mathrm{s}}^{{H}}{\mathbf{P}}_{\mathrm{d}}^ \bot } \right)} \right\}\\
&= \min \left\{ {{\operatorname{rank}}\left( {{{\mathbf{\Phi }}_{\mathrm{s}}}} \right),{\operatorname{rank}}\left( {{\mathbf{\Phi }}_{\mathrm{s}}^{{H}}} \right)} \right\} = {\operatorname{rank}}\left( {{{\mathbf{\Phi }}_{\mathrm{s}}}} \right).
\end{aligned}
\end{equation}

From Lemma 2, the noncentrality parameters $\left( {{\delta }_{0}},{{\delta }_{1}} \right)$ are respectively expressed as
\begin{equation}
\begin{aligned}
{\delta _0} &= {\left( {\frac{{{{\mathbf{\Phi }}_{\mathrm{d}}}{\mathbf{a}}}}{{\sqrt N \sigma }}} \right)^{{H}}}{\mathbf{P}}_{\mathrm{d}}^ \bot {{\mathbf{\Phi }}_{\mathrm{s}}}{\left[ {{\mathbf{\Phi }}_{\mathrm{s}}^{{H}}{\mathbf{P}}_{\mathrm{d}}^ \bot {{\mathbf{\Phi }}_{\mathrm{s}}}} \right]^{ - 1}}{\mathbf{\Phi }}_{\mathrm{s}}^{{H}}{\bf{P}}_{\mathrm{d}}^ \bot \left( {\frac{{{{\mathbf{\Phi }}_{\mathrm{d}}}{\mathbf{a}}}}{{\sqrt N \sigma }}} \right)\\
{\delta _1} &= {\left( {\frac{{{{\mathbf{\Phi }}_{\mathrm{d}}}{\mathbf{a}} + {{\mathbf{\Phi }}_{\mathrm{s}}}{\mathbf{b}}}}{{\sqrt N \sigma }}} \right)^{{H}}}{\mathbf{P}}_{\mathrm{d}}^ \bot {{\mathbf{\Phi }}_{\mathrm{s}}}{\left[ {{\mathbf{\Phi }}_{\mathrm{s}}^{{H}}{\mathbf{P}}_{\mathrm{d}}^ \bot {{\mathbf{\Phi }}_{\mathrm{s}}}} \right]^{ - 1}}\\
&\cdot {\mathbf{\Phi }}_{\mathrm{s}}^{{H}}{\mathbf{P}}_{\mathrm{d}}^ \bot \left( {\frac{{{{\mathbf{\Phi }}_{\mathrm{d}}}{\mathbf{a}} + {{\mathbf{\Phi }}_{\mathrm{s}}}{\mathbf{b}}}}{{\sqrt N \sigma }}} \right).
\end{aligned}
\end{equation}

Thus, the probability of false alarm and detection can be written as 
\begin{equation}\label{28}
\begin{aligned}
{P_{FA}} &= \Pr \left\{ {{T_0}({\mathbf{X}}) > \eta ;{{\cal H}_0}} \right\} = {Q_{\chi _v^2({\delta _0})}}(\eta ),\\
{P_D} &= \Pr \left\{ {{T_0}({\mathbf{X}}) > \eta ;{{\cal H}_{\rm{1}}}} \right\} = {Q_{\chi _v^2({\delta _1})}}(\eta ).
\end{aligned}
\end{equation}

The right tail probability for noncentral central chi-squared distribution can be achieved using the recursive method proposed by Mitchell and Walker \cite{mitchell1971recursive}.

\subsection{Unknown spectrum power of noise}
GLR can be expressed as
\begin{equation}\label{29}
\begin{split}
\begin{aligned}
&{L_{{\mathrm{GLR}}}}({\mathbf{X}}) = \frac{{{{\max }_{{\mathbf{a}},{\mathbf{b}},{{\mathbf{\tau }}^\mathrm{d}},{{\mathbf{\tau }}^{\mathbf{s}}},{\sigma ^2}}}f\left( {{\mathbf{X}}|{\mathbf{a}},{\mathbf{b}},{{\mathbf{\tau }}^{\mathrm{d}}},{{\mathbf{\tau }}^{\mathrm{s}}},{\sigma ^2};{{\cal H}_1}} \right)}}{{{{\max }_{{\mathbf{a}},{{\mathbf{\tau }}^{\mathrm{d}}},{\sigma ^2}}}f\left( {{\mathbf{X}}|{\mathbf{a}},{{\mathbf{\tau }}^{\mathrm{d}}},{\sigma ^2};{{\cal H}_0}} \right)}} \\
&= {\left( {\frac{{\sigma _0^2}}{{\sigma _1^2}}} \right)^{N/2}}\\
&= {\left( {\frac{{{{[{\mathbf{X}} - {{\mathbf{\Phi }}_{\mathrm{d}}}{{\widehat {\mathbf{a}}}_0}]}^{H}}[{\mathbf{X}} - {{\mathbf{\Phi }}_{\mathrm{d}}}{{\widehat {\mathbf{a}}}_0}]}}{{{{[{\mathbf{X}} - {{\mathbf{\Phi }}_{\mathrm{d}}}{{\widehat {\mathbf{a}}}_{\mathrm{1}}} - {{\mathbf{\Phi }}_{\mathrm{s}}}{{\widehat {\mathbf{b}}}_{\mathrm{1}}}]}^{H}}[{\mathbf{X}} - {{\mathbf{\Phi }}_{\mathrm{d}}}{{\widehat {\mathbf{a}}}_{\mathrm{1}}} - {{\mathbf{\Phi }}_{\mathrm{s}}}{{\widehat {\mathbf{b}}}_{\mathrm{1}}}]}}} \right)^{N/2}}\\
&= {\left( {\frac{{{{\mathbf{X}}^H}({\mathbf{I}} - {{\mathbf{\Phi }}_{\mathrm{d}}}{{({\mathbf{\Phi }}_{\mathrm{d}}^{H}{{\mathbf{\Phi }}_{\mathrm{d}}})}^{ - {\mathbf{1}}}}{\mathbf{\Phi }}_{\mathrm{d}}^{H}){\mathbf{X}}}}{{{{\mathbf{X}}^H}({\mathbf{I}} - {\mathbf{Q}}{{({{\mathbf{Q}}^{H}}{\mathbf{Q}})}^{ - {\mathbf{1}}}}{{\mathbf{Q}}^{H}}){\mathbf{X}}}}} \right)^{N/2}}.
\end{aligned}
\end{split}
\end{equation}

Let the test statistic be
\begin{equation}
\begin{aligned}
{T_1}({\mathbf{X}}) &= {L_{{\mathrm{GLR}}}}{({\mathbf{X}})^{2/N}} - 1\\
&= \frac{{{{\mathbf{X}}^{H}}{\mathbf{P}}_{\mathrm{d}}^ \bot {{\mathbf{\Phi }}_{\mathrm{s}}}{{\left[ {{\mathbf{\Phi }}_{\mathrm{s}}^{H}{\mathbf{P}}_{\mathrm{d}}^ \bot {{\mathbf{\Phi }}_{\mathrm{s}}}} \right]}^{ - 1}}{\mathbf{\Phi }}_{\mathrm{s}}^{H}{\mathbf{P}}_{\mathrm{d}}^ \bot {\mathbf{X}}}}{{{{\mathbf{X}}^{H}}{\mathbf{P}}_{\mathrm{d}}^ \bot {\mathbf{X}}}}\\
&= \frac{{{{\mathbf{X}}^{H}}{\mathbf{P}}_{\mathrm{d}}^ \bot {{\mathbf{\Phi }}_{\mathrm{s}}}{{\left[ {{\mathbf{\Phi }}_{\mathrm{s}}^{H}{\mathbf{P}}_{\mathrm{d}}^ \bot {{\mathbf{\Phi }}_{\mathrm{s}}}} \right]}^{ - 1}}{\mathbf{\Phi }}_{\mathrm{s}}^{H}{\mathbf{P}}_{\mathrm{d}}^ \bot {\mathbf{X}}{\mathrm{/(}}N{\sigma ^2})}}{{{{\mathbf{X}}^{H}}{\mathbf{P}}_{\mathrm{d}}^ \bot {\mathbf{X}}{\mathrm{/(}}N{\sigma ^2})}}\\
&= \frac{{M({\mathbf{X}})}}{{D({\mathbf{X}})}}
\mathop{\gtrless}\limits_{{\cal H}_0}^{{\cal H}_1} \eta.
\end{aligned}
\end{equation}

Similar to \eqref{25}, we obtain
\begin{equation}\label{31}
\begin{aligned}
M({\mathbf{X}})\sim\left\{ {\begin{array}{*{20}{c}}
	{\chi _v^2({\delta _0})}&{{\text{under }}{{\cal H}_0}}\\
	{\chi _v^2({\delta _1})}&{{\text{under }}{{\cal H}_1}}
	\end{array}} \right.,
\end{aligned}
\end{equation}
\begin{equation}
\begin{aligned}
D({\mathbf{X}})\sim\left\{ {\begin{array}{*{20}{c}}
	{\chi _r^2({\lambda _0})}&{{\text{under }}{{\cal H}_0}}\\
	{\chi _r^2({\lambda _1})}&{{\text{under }}{{\cal H}_1}}
	\end{array}} \right.,
\end{aligned}
\end{equation}
where $r$ is the rank of matrix $\mathbf{P}_{\text{d}}^{\bot }$. The noncentrality parameters are given by
\begin{equation}
\begin{aligned}
{\lambda _0} &= {\left( {\frac{{{{\mathbf{\Phi }}_{\mathrm{d}}}{\mathbf{a}}}}{{\sqrt N \sigma }}} \right)^{H}}{\mathbf{P}}_{\mathrm{d}}^ \bot \left( {\frac{{{{\mathbf{\Phi }}_{\mathrm{d}}}{\mathbf{a}}}}{{\sqrt N \sigma }}} \right),\\
{\lambda _1} &= {\left( {\frac{{{{\mathbf{\Phi }}_{\mathrm{d}}}{\mathbf{a}} + {{\mathbf{\Phi }}_{\mathrm{s}}}{\mathbf{b}}}}{{\sqrt N \sigma }}} \right)^{H}}{\mathbf{P}}_{\mathrm{d}}^ \bot \left( {\frac{{{{\mathbf{\Phi }}_{\mathrm{d}}}{\mathbf{a}} + {{\mathbf{\Phi }}_{\mathrm{s}}}{\mathbf{b}}}}{{\sqrt N \sigma }}} \right).
\end{aligned}
\end{equation}

By normalizing the numerator and denominator with degrees of freedom, we achieve a new equivalence test statistic 
\begin{equation}
{T_1}'({\mathbf{X}}) = \frac{r}{v}{T_1}({\mathbf{X}}) = 
\frac{{M({\mathbf{X}}){/}v}}{{D({\mathbf{X}}){/}}r}
\mathop{\gtrless}\limits_{{\cal H}_0}^{{\cal H}_1} \eta.
\end{equation}

Clearly, the distribution of the test statistic is doubly noncentral F distribution, which is expressed as
\begin{equation}
\begin{aligned}
{T_1}'({\mathbf{X}})\sim\left\{ {\begin{array}{*{20}{c}}
	{F(v,r,{\delta _0},{\lambda _0})}&{{\text{under }}{{\cal H}_0}}\\
	{F(v,r,{\delta _1},{\lambda _1})}&{{\text{under }}{{\cal H}_{\rm{1}}}}
	\end{array}} \right..
\end{aligned}
\end{equation}

Thus, the probability of false alarm and detection can be written as
\begin{equation}\label{36}
\begin{aligned}
{P_{FA}} = \Pr \left\{ {{T_1}'({\mathbf{X}}) > \eta ;{{\cal H}_0}} \right\} = {Q_{F(v,r,{\delta _0},{\lambda _0})}}(\eta ),\\
{P_D} = \Pr \left\{ {{T_1}'({\mathbf{X}}) > \eta ;{{\cal H}_1}} \right\} = {Q_{F(v,r,{\delta _1},{\lambda _1})}}(\eta ).
\end{aligned}
\end{equation}

The right tail probability for the doubly noncentral F distribution can be calculated using the method proposed by Marc S. Paolella \cite{paolella2007intermediate}.

\section{Simulation results}
\subsection{Experimental configuration}
\begin{figure}[ht]
	
	\centering
	\includegraphics[scale=0.3]{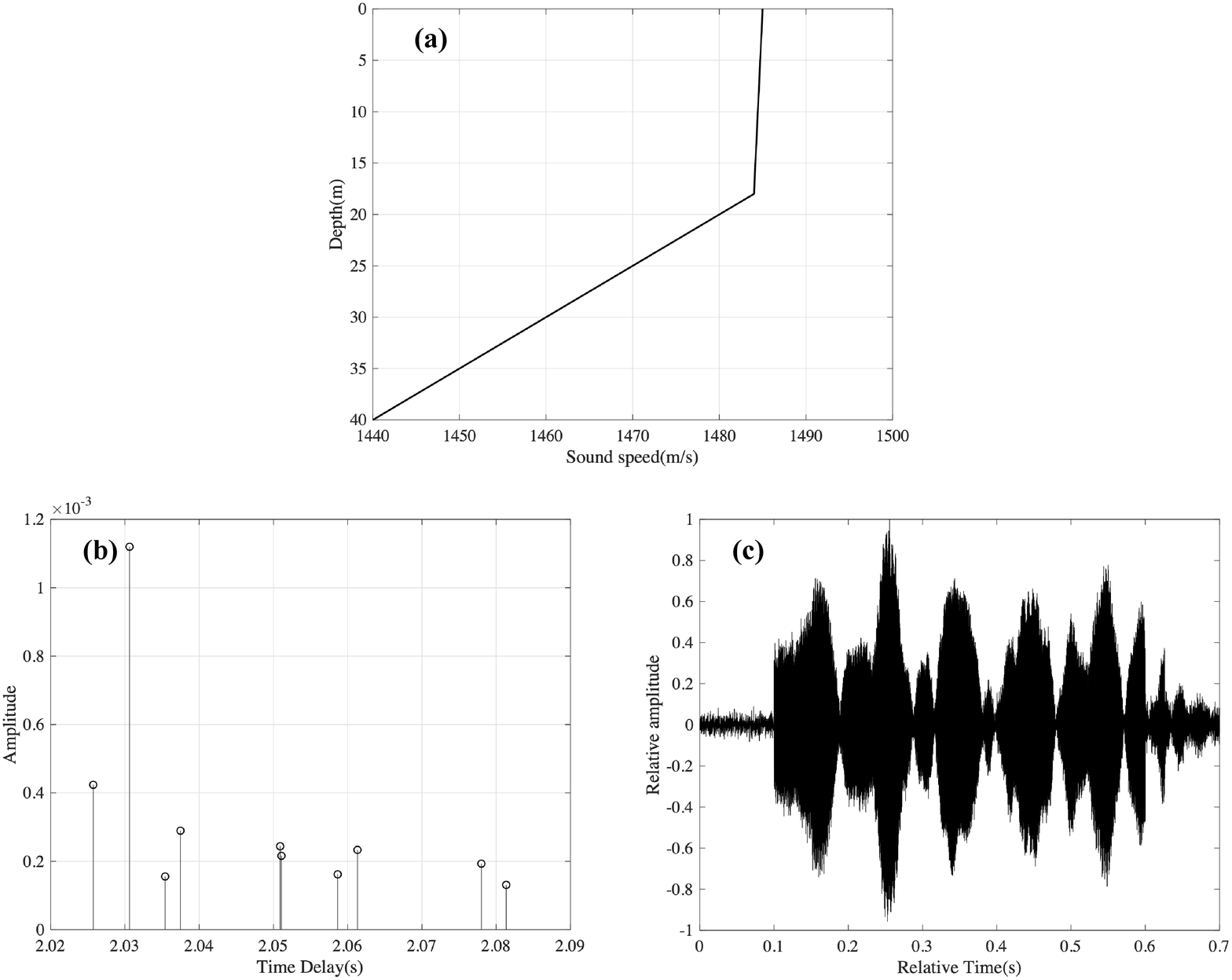}
	\caption{Acoustic environment and received waveforms. (a) Sound speed profile in shallow water. The upper volume is approximate isovelocity, and the lower volume has a negative gradient sound speed profile. (b) Channel impulse response simulated using the BELLHOP ray model. (c) Received signal without the target using the BELLHOP ray model.}
	\label{fig2}
\end{figure}

In a shallow water environment, the bistatic sonar consists of a transmitter and a receiver separated by 3 km and located at a depth of 10 m. The water depth is set to 40 m, and the sound speed profile is shown in Fig. \ref{fig2}(a). The bottom is assumed to be a half-space with a density of 1.6 g/cm$^{3}$ and sound speed of 1720 m/s. A linear frequency modulation (LFM) signal with a duration of 0.5 s, center frequency of 2 kHz, and bandwidth of 200 Hz is transmitted repeatedly every 2 s. The received signals are simulated using the BELLHOP ray model \cite{porter1987gaussian} with the consideration of 10 main propagation paths shown in Fig. \ref{fig2}(b). The received signal is shown in Fig. \ref{fig2}(c). The intruder is a rigid prolate ellipsoid with a length of 8 and 2 m for the long and short halves, respectively.

\begin{figure}[ht]
	
	\centering
	\includegraphics[scale=0.3]{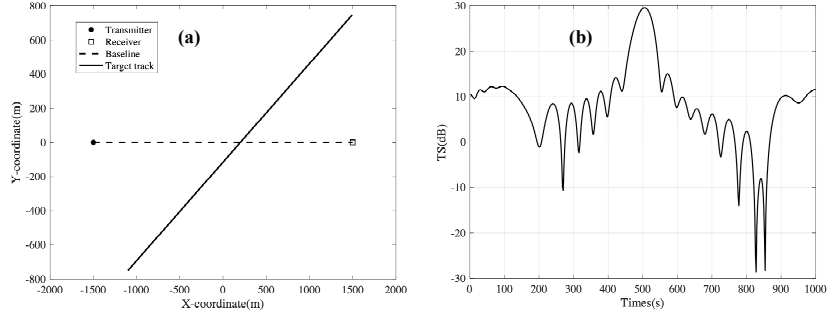}
	\caption{Target position and target strength during the crossing event. (a) Bistatic configuration and target trajectory. (b) Corresponding target strength under bistatic configuration.}
	\label{fig3}
\end{figure}

\begin{figure}[ht]
	
	\centering
	\includegraphics[scale=0.3]{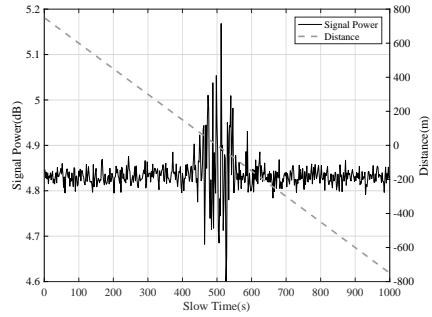}
	\caption{Variation in the received signal power. When the target reaches the baseline, the maximum fluctuation occurs, but it does not exceed 0.5 dB.}
	\label{fig4}
\end{figure}

To simplify the problem, we assume that the target is at the same depth as the transceiver and that the target pitch angle is 0 degree. As shown in Fig. \ref{fig3}(a), the underwater target travels straight through the source-receiver line at a speed of 3 m/s in an observation duration of 1000 s. At 500 s, the target crosses the source-receiver line at (200, 0)m on the horizontal plane. Given that the movement of the target changes the geometric relationship between the target and bistatic system, the target strength (TS) fluctuates continuously during the closing event. The bistatic separation angle increases as the target approaches the baseline. It is 180° when the target is on the baseline. The bistatic separation angle gradually decreases during the departure event. Fig. \ref{fig3}(b) shows the TS variation for the bistatic system during the crossing event calculated using the deformed cylinder method \cite{ye1997method}. During this crossing event, the receiver receives a total of 500 sets of transmitted signal pulses, and the sampling rate is set to 10 kHz. In the event of the intruder crossing the source-receiver line, an aberration in the received signals occurs due to the interference between the direct blast and scattered signal. When an object crosses at midpoint between the source and receiver, the acoustic field aberration is at the minimum \cite{lei2012forward}. Hence, to show the performance of the method in forward detection, a worse situation is considered in the simulation. Multipath time delays are estimated under a signal-to-noise ratio (SNR) of 0 dB in advance (the scattered signal strength constantly varies with the movement, so the scattered signals at the crossing point are used as a reference, and the signal-to-direct blast ratio (SDR) at this moment is approximately -18.5 dB).As shown in Fig. \ref{fig4}, Gaussian white noise is added to the received signals with an SNR of 0 dB. When the target reaches the baseline, the maximum fluctuation occurs, but it does not exceed 0.5 dB.

\subsection{Detection Results}

\begin{figure}[ht]
	
	\centering
	\includegraphics[scale=0.3]{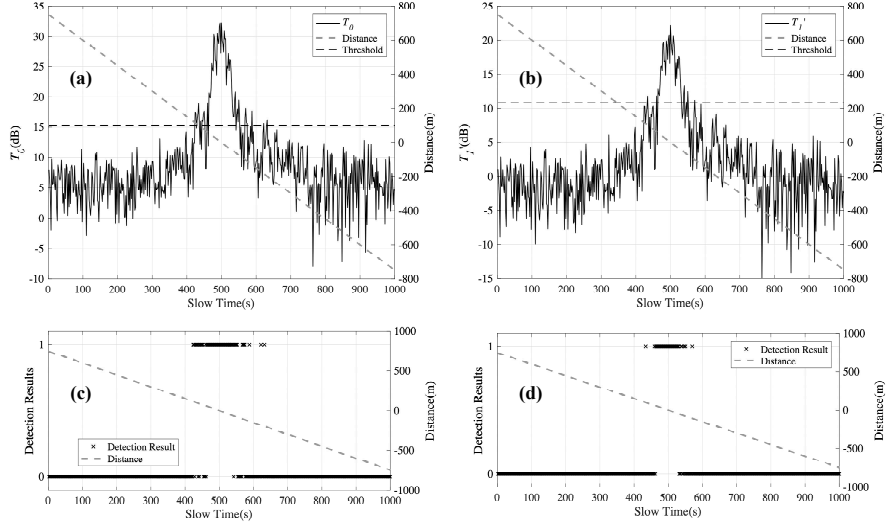}
	\caption{GLRT results in the crossing event. (a) and (b) are the GLR of each received signal. The dashed line indicates the distance from the target to the baseline. (c) and (d) are the detection result for   and   with  , respectively. 1 indicates that the target is detected, and 0 indicates that no target is present.}
	\label{fig5}
\end{figure}

GLRT for known $\sigma^2$ and unknown $\sigma^2$ are performed on each of the received signals, and the detection results are shown in Figs. \ref{fig5}(a) and \ref{fig5}(b), respectively. The received signal achieves a high GLR when the target is beyond the baseline. The detection results are shown in Figs. \ref{fig5}(c) and \ref{fig5}(d), with the threshold set at 15.28 and 10.79 dB from \eqref{28} and \eqref{36}, respectively, under the probability of false alarm ${{10}^{-6}}$. Given that the estimated multipath delays will not exactly match the true values when the target is moving, some missing detection exists in the detection results. Moreover, ${{T}_{0}}$ performs better than ${{T}_{1}}'$ because it has more prior information on noise power.

\subsection{Influence of Noises}
\begin{figure}[ht]
	
	\centering
	\includegraphics[scale=0.3]{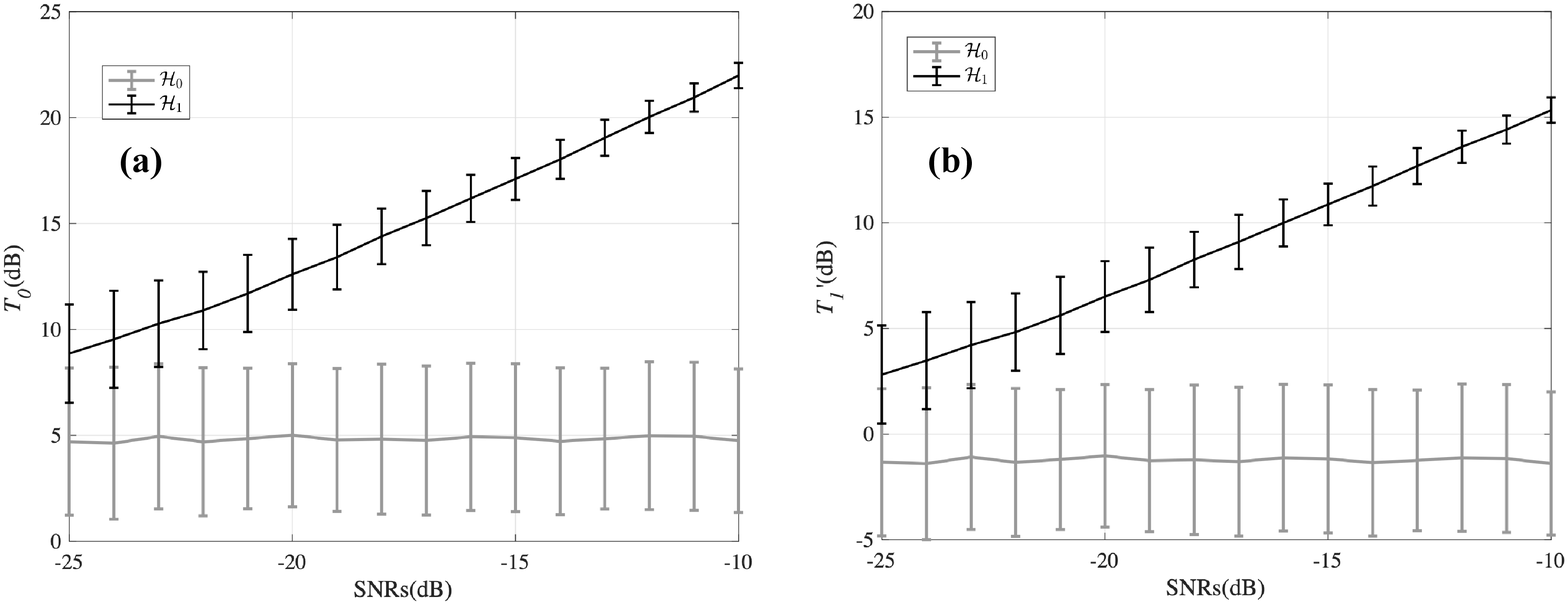}
	\caption{${{T}_{0}}$ of hypothesis $\mathcal{H}_{0}$ and ${{T}_{1}}'$ of hypothesis $\mathcal{H}_{1}$ under different SNRs. The error bars indicate standard deviation.}
	\label{fig6}
\end{figure}

\begin{figure}[ht]
	
	\centering
	\includegraphics[scale=0.3]{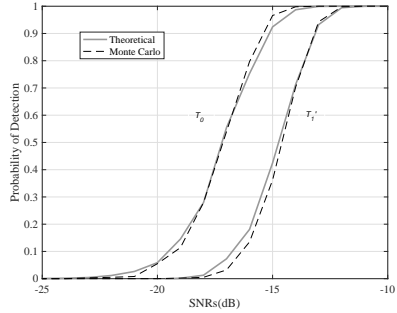}
	\caption{Probability of detection under different SNRs for ${{P}_{FA}={10}^{-6}}$.}
	\label{fig7}
\end{figure}

Further analysis of the method is performed. We add Gaussian white noise to obtain different SNRs. The GLRs of hypotheses $\mathcal{H}_{0}$ and $\mathcal{H}_{1}$ computed with 1000 Monte Carlo simulations are shown in Fig. \ref{fig6}. As SNR decreases, the value of GRL under hypothesis $\mathcal{H}_{1}$ approaches that under hypothesis $\mathcal{H}_{0}$, and their variances increase. With the setting ${{P}_{FA}}={{10}^{-6}}$, the Monte Carlo experimental results are consistent with the theoretical values obtained by \eqref{28} and \eqref{36}, as shown in Fig. \ref{fig7}. Minimal variations originate from the error of computation. Furthermore, the performance of ${{T}_{0}}$ is much better than that of ${{T}_{1}}'$. Under the SNR of -15dB, the probability of detection is greater than 0.9 for ${{T}_{0}}$ but is under 0.5 for ${{T}_{1}}'$. When a probability of detection of 0.9 is required, SNRs of approximately -15 and -13 dB are needed for ${{T}_{0}}$ and ${{T}_{1}}'$, respectively. This condition indicates that the performance decreases by 2-3 dB when the noise power is unknown.

\subsection{Influence of SDR}
\begin{figure}[ht]
	
	\centering
	\includegraphics[scale=0.3]{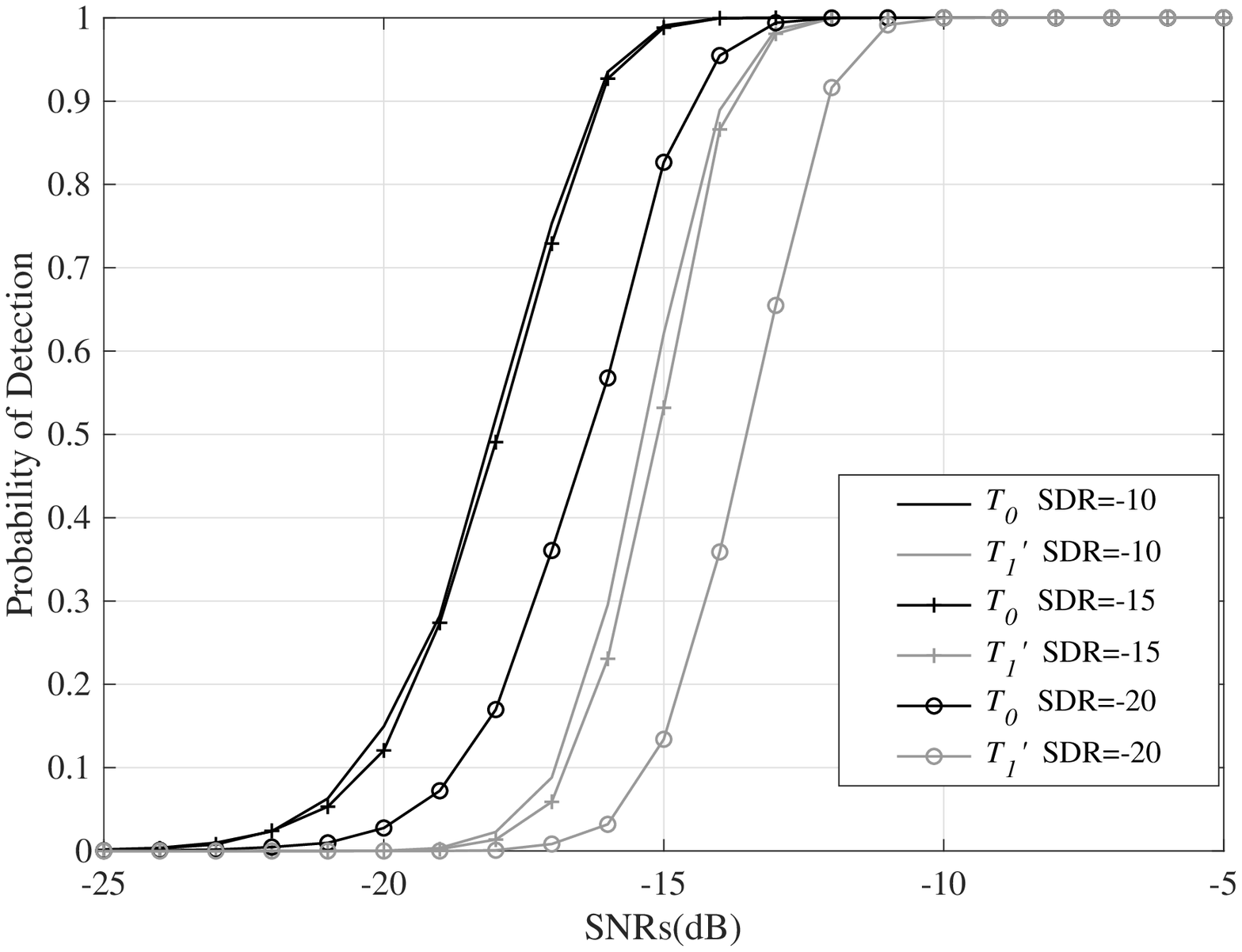}
	\caption{Probability of detection under different SDRs for ${{P}_{FA}={10}^{-6}}$.}
	\label{fig8}
\end{figure}

As can be seen from \eqref{25} and \eqref{31}, detection performance has a significant relationship with the direct blast, which depends on ${{\Phi }_{\text{d}}}$. Thus, the performance is associated with SDR. In this part, the performance of the detection method is analyzed in consideration of SDR. The results under different SDRs for ${{T}_{0}}$ and ${{T}_{1}}'$ are shown in Fig. \ref{fig8}. A large SDR has better detection performance than a low SDR. When the SDR is -15 dB, a small decrease in performance occurs compared with the SDR of -10 dB. When the SDR decreases to -20 dB, a large decrease occurs. The probability of detection decreases sharply when SDR decreases.

\subsection{Influence of Number of Paths}
\begin{figure}[ht]
	
	\centering
	\includegraphics[scale=0.3]{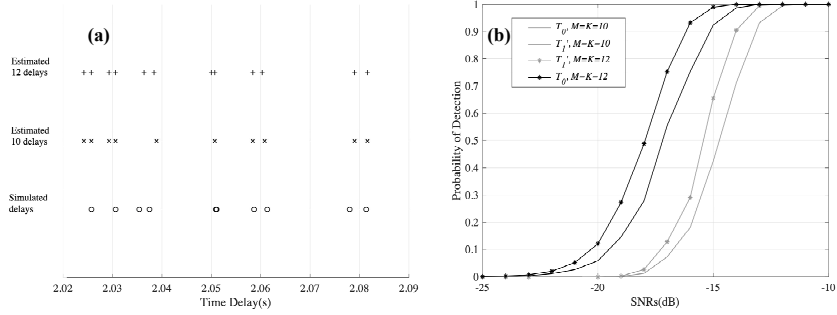}
	\caption{Influence of the number of paths. (a) Comparison of time delay estimation results. (b) Probability of detection under different numbers of paths for ${{P}_{FA}}={{10}^{-6}}$}
	\label{fig9}
\end{figure}

As discussed in the previous parts, time delay estimates exert a great impact on detection performance. Time delay resolution is associated with the frequency band of the transmitted signal. Generally, a wideband signal provides a high resolution in time delays, such that time delay estimates may be accurately achieved with super-resolution methods \cite{ge2007super,yun2014improved}. In the time delay estimation procedure, two similar estimated values may appear sometimes, particularly in cases with low SNRs. When the number of estimated paths is more than the number of main paths in reality, the delay estimates are well evaluated even though some of the paths are duplicate. As shown in Fig. \ref{fig9}(a), duplicate estimates exist when 10 paths are considered. Meanwhile, the four delay estimates of about 2.04 and 2.05 are accurate when 12 paths are considered. This finding indicates that this method works when the number of estimated paths reaches or exceeds the number of main paths of the received signal. As shown in Fig. \ref{fig9}(b), when there are 10 paths for the received signal, the method works if we make 10 multipath delay estimations. A slight improvement in the method is observed when 12 multipath delays are estimated. For ${{T}_{0}}$, the performance is about 1 dB for SNRs and not more than 1 dB for ${{T}_{1}}'$. Notably, when the number of estimated paths is smaller than the number of main paths of the received signal, the distributions of the test statistics under the two hypotheses are almost the same, resulting in the failure of the method. If the number of estimated paths is much larger than the number of true paths, then the inversion of matrix in \eqref{24} and \eqref{29} will be singular.

\subsection{Influence of Number of FFT Points}
\begin{figure}[ht]
	
	\centering
	\includegraphics[scale=0.3]{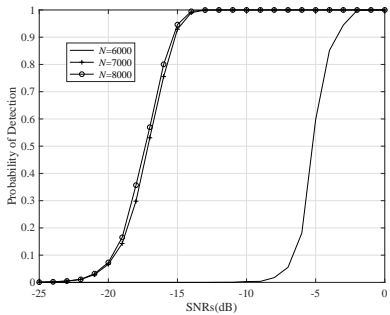}
	\caption{Probability of detection under different FFT lengths for ${{P}_{FA}}={{10}^{-6}}$.}
	\label{fig10}
\end{figure}

Only the effect of the number of FFT points $N$ on ${{T}_{0}}$ is calculated here because numerical calculation of the right-tailed probability of the double non-center F distribution has a low degree of confidence under high degrees of freedom. The length of the signal in the time domain is 7000. When the number of FFT points is smaller than the signal length, truncation occurs and leads to a considerable reduction in performance. Meanwhile, increasing the number of points to above 8000 will improve the performance of this method slightly, as shown in Fig. \ref{fig10}. Specifically, calculating ${{T}_{0}}$ requires ${{N}^{3}}+(3M+2){{N}^{2}}+6{{M}^{2}}N$ multiplications, and calculating ${{T}_{1}}'$ requires${{N}^{3}}+(3M+4){{N}^{2}}+6{{M}^{2}}N$ multiplications. Hence, in-creasing the number of points $N$ will improve the computation burden with a cubic in-crease. 

\section{Conclusion}
A strong direct blast restricts the application of bistatic sonar and makes the area near the baseline a blind zone for detection. This study presents a novel and robust method that addresses this problem. 

On the basis of the generalized likelihood ratio test, the corresponding test statistics for known and unknown noise power are proposed by building a frequency domain received signal model. The noncentral central chi-squared distribution and doubly noncentral F distribution for known and unknown noise power situations, respectively, are given in theory. The intrinsic CFA is thus represented. 

The developed method requires the transmitted signal to have a certain duration because multipath propagation overlap in the receiver is required to satisfy the frequency domain model in hypothesis. In addition, the method is sensitive to multipath delay estimations; hence, the transmitted signal needs bandwidth. Although a high-resolution time delay estimation method may be used, a large bandwidth is still beneficial for accurate time delay estimation.

The simulation results reveal the effectiveness of the method under forward scattering detection configuration with a strong blast. The sensitivity of many factors, such as noise, SDR, number of paths, and FFT size, is discussed. The theoretical and simulated results seem promising, and further sea trials are needed.

\section{Lemma involved in the article}
\subsection{Lemma 1}
For a partitioned matrix, if all inverses exist for proper dimensions of the matrices, then
\begin{equation}
{\left[ {\begin{array}{*{20}{c}}
		{\mathbf{A}}&{\mathbf{B}}\\
		{\mathbf{E}}&{\mathbf{F}}
		\end{array}} \right]^{ - 1}} = \left[ {\begin{array}{*{20}{c}}
	{{{\mathbf{A}}^{ - 1}} + {{\mathbf{A}}^{ - 1}}{\mathbf{BKA}}}&{ - {{\mathbf{A}}^{ - 1}}{\mathbf{BK}}}\\
	{ - {\mathbf{KC}}{{\mathbf{A}}^{ - 1}}}&{\mathbf{K}}
	\end{array}} \right],
\end{equation}
\begin{equation}
{\mathbf{K}} = {\left( {{\mathbf{F}} - {\mathbf{E}}{{\mathbf{A}}^{ - 1}}{\mathbf{B}}} \right)^{ - 1}}.
\end{equation}

\subsection{Lemma 2}
Let $\mathbf{X} \sim C N(u, \mathbf{I})$ be a complex $N\times1$ vector, and let $\mathbf{A}$ be a Hermitian matrix of dimension $N\times N$ . Then, $\mathbf{X}^{H} \mathbf{A} \mathbf{X}$ has a noncentral, complex, chi-squared random variable with $k$ complex degrees of freedom and noncentrality parameter $\delta$ if and only if $\mathbf{A}$ is an idempotent matrix, in which case the complex degree of freedom and noncentrality parameters are $k=\operatorname{rank}(\mathbf{A})=\operatorname{tr}(\mathbf{A})$ and $\delta=\mu^{H} \mathbf{A} \mu$, respectively.

\bibliographystyle{IEEEtran}
\bibliography{IEEEabrv,IEEEexample}

\end{document}